\documentclass[aps,pre,preprint,showpacs,superscriptaddress]{revtex4}
\usepackage{psfrag}
\usepackage{amssymb,amsmath,amsthm}
\usepackage[dvips]{graphicx,color,epsfig}
\usepackage{verbatim}

\begin{document}
\title{STATISTICAL COMPLEXITY AND NONTRIVIAL COLLECTIVE BEHAVIOR IN ELECTROENCEPHALOGRAPHIC SIGNALS}
\author{M. ESCALONA-MOR\'AN}
\affiliation{Grupo de Ingenier\'ia Biom\'edica, Facultad de Ingenier\'ia, Universidad de Los Andes, 
M\'erida, Venezuela.}
\affiliation{Laboratoire Traitement du Signal et de l'Image (LTSI), Universit\'e de Rennes 1, 
Campus Scientifique de Beaulieu, B\^at. 22, 35042 Rennes Cedex, France.}
\affiliation{INSERM U642, LTSI, B\^at. 22, 35042 Rennes Cedex, France}
\author{M. G. COSENZA}
\affiliation{Centro de F\'isica Fundamental, Universidad de Los Andes, M\'erida, Venezuela.}
\author{R. L\'OPEZ-RUIZ}
\affiliation{DIIS and BIFI, Facultad de Ciencias, 
Universidad de Zaragoza, E-50009 Zaragoza, Spain.}
\author{P. GARC\'IA}
\affiliation{Laboratorio de Sistemas Complejos,  Departamento de F\'isica Aplicada,  
Facultad de Ingenier\'ia, Universidad Central de Venezuela, Caracas, Venezuela.}

\begin{abstract}
We calculate a measure of statistical complexity from the global dynamics of electroencephalographic (EEG) 
signals from healthy subjects and epileptic patients, and are able to establish a criterion to characterize 
the collective behavior in both groups of individuals. It is found that the collective dynamics of EEG signals 
possess relative higher values of complexity for healthy subjects in comparison to that for epileptic patients. 
To interpret these results, we propose a model of a network of coupled chaotic maps where we calculate the 
complexity as a function of a parameter and relate this measure with the emergence of nontrivial collective 
behavior in the system. Our results show that the presence of nontrivial collective behavior is associated 
to high values of complexity; thus suggesting that similar dynamical collective process may take place in the 
human brain. Our findings also suggest that epilepsy is a degenerative illness related to the loss of 
complexity in the brain.
\end{abstract}

\pacs{05.45.-a, 05.45.Xt, 05.45.Ra}
\maketitle

The concept of complex systems has become a new paradigm for the search of mechanisms
and a unified interpretation of
the  processes of emergence of structures, organization and functionality
in a variety of natural
and artificial phenomena in different contexts
[Badii \& Politi, 1997; Mikhailov \& Calenbuhr, 2002;  Kaneko \& Tsuda, 2000].
One common criterion for defining complexity is the emergent
behavior: collective structures, patterns and functions that are absent 
at the local level arise from
simple interaction rules between the constitutive elements in a system.
Phenomena such as the spontaneous formation of structures, organization,
spatial patterns, chaos synchronization, collective
oscillations, spiral waves, segregation and differentiation, formation and
growth of domains, and social consensus, are examples
of self-organizing processes that occur in various contexts such as
physical, chemical, biological, physiological, social and economic systems.

There has been much interest in the study of
the phenomenon of emergence of nontrivial collective behavior in the context of
systems of interacting chaotic elements 
[Kaneko, 1990; Chat\'e \& Manneville, 1992a, 1992b; Pikovsky \& Kurths, 1994; Shibata \& Kaneko, 1998; 
Cosenza, 1998; Cosenza \& Gonz\'alez, 1998; Cisneros \textit{et al.}, 2002; Manrubia \textit{et al.}, 2004].
Nontrivial collective behavior is characterized by a well defined
evolution of macroscopic quantities coexisting with 
local chaos. Models based on coupled map networks have
been widely used in the investigation of collective phenomena
that appear in many complex systems [Kaneko \& Tsuda, 2000]. In particular, networks of coupled chaotic maps can 
exhibit nontrivial collective behavior.

A paradigmatic example of a complex system is provided by the human brain.
It consists of a highly interconnected network of millions of neurons. The local 
dynamics of a neuron in general behaves as a non-linear excitable element [Herz \textit{et al.}, 2006]. 
From the signal of a single neuron it is not 
possible to understand the highly structured collective behavior and 
functions of the brain.

In this paper we investigate the relative complexity of the human brain by considering the 
collective dynamics that arise from the local dynamics of groups of neurons, as manifested in 
electroencephalographic (EEG) signals. We calculate a measure of
complexity from the global dynamics of EEG signals from healthy subjects and epileptic patients, 
and are able to establish a criterion to characterize the collective behavior in both groups of 
individuals. It is found that the collective dynamics of EEG signals possess relative higher values 
of complexity for healthy subjects in comparison to that for epileptic patients. 
Our results support the view that epilepsy is characterized by a
loss of complexity in the brain, as indicated by measurements of the dimension correlation [Babloyantz A. \& Destexhe A., 1986],
algorithmic complexity [Rapp \textit{et al.}, 1994], and anticipation
of seizures [Martinerie  \textit{et al.}, 1998]. 

In order to interpret our results, we propose a model of coupled chaotic maps where we calculate the measure of complexity 
as a function of a parameter and relate this measure with the emergence of nontrivial collective behavior 
in the system. Our results show that the appearance of nontrivial collective behavior is associated 
to high values of complexity; thus suggesting that similar dynamical collective process may take 
place in the human brain.  

Several measures of complexity have been proposed in the literature. Here
we employ the concept of statistical complexity, introduced by Lopez-Ruiz \textit{et al.} [1995]. 
This quantity is based on the statistical description of a system at a given 
scale, and it has been shown to be capable of discerning among different 
macroscopic structures emerging in complex 
systems [S\'anchez \& Lopez-Ruiz, 2005].
The amount of complexity $C$ is obtained by 
computing the product between the entropy $H$, and a sort of distance to 
the equipartition state in the system, named the disequilibrium $D$. Thus, the statistical complexity 
is defined as [Lopez-Ruiz \textit{et al.}, 1995]
\begin{equation}
\label{complexityEq}
C = H \cdot D = -K\sum_{s=1}^{\mathcal{R}} p_s \log p_s \cdot 
\sum_{s=1}^{\mathcal{R}} \Big(p_s - \frac{1}{\mathcal{R}}\Big)^2,
\end{equation}
where $H$ and $D$ are, respectively, the entropy and the disequilibrium; 
$p_s$ represents the probability associated to the state $s$;  $\mathcal{R}$ is
the number of states, 
and $K$ is a positive normalization constant. Note that $p_s$ may vary for different
levels of observation, reflected in 
$\mathcal{R}$. The quantity $C$ can quantify relative values of complexity in a specific system at 
a given level of description.

The EEG data base used in this study consists of records from 
$40$ individuals in an age range between $22$ and $48$ years old. These individuals are classified into four groups:  
(I) a group of $10$ healthy subjects; (II) a group of $10$ epileptic patients receiving treatment with Phenobarbital for at least $18$ months; (III) a group consisting of $18$ epileptic patients that have not yet received medical treatment; and (IV) a group
of $2$ epileptic patients who experienced spontaneous seizures during the EEG recording.
All the epileptic patients in the data base were diagnosed generalized epilepsy manifested through tonic-clonic seizures.

The record of the EEG signal from each individual was carried out over $19$ channels connected 
to scalp electrodes according to the international $10-20$ system [Jasper, 1958]. 
The potentials were measured with respect to a reference level 
consisting of both ears short-circuited. The signal was digitalized at a sampling frequency of $256$~Hz and 
A/D conversion of $12$ bits, and filtered to bandwidth between $0.5$ and $30$~Hz. All the EEG signals were recorded with the individuals at rest and with eyes closed, during a one-hour period, between $8$~a.m. and $10$~a.m.  
We consider continuous segments of the EEG signals containing between $15000$ to $30000$ points, with no significant artifacts. 

A channel in the EEG signal represents an average of a set of neurons in an specific part of the brain.
The $19$ channels in each EEG can be considered as simultaneous time series coming from different parts of 
an interacting dynamical network. 
The collective behavior of such a dynamical network at a given time $t$ can be described by, 
the instantaneous mean field $h_t$, defined as
\begin{equation}
\label{meanFieldeegEq}
h_t = \frac{1}{M}\sum_{j=1}^M e_t^{j},
\end{equation}
where $e_t^j$ is the real value registered by electrode $j$ at discrete time $t$, $j=1,\ldots,M$; 
and $M=19$ is the number of electrodes. 

The probability distribution of the mean field values corresponding to a given EEG signal is constructed 
from the time series of $h_t$ calculated for that signal. We define the number of states $\mathcal{R}$ as
a partition consisting of $\mathcal{R}$ equal size segments on the 
range of values of $h_t$. Next, the probability $p_s$ associated to the $s$ state is calculated. Here we set $\mathcal{R}=10^3$ for all the EEG signals. Then, Eq.~(\ref{complexityEq}) is used to calculate the statistical complexity $C$ associated to the mean field for each EEG.

\begin{figure}[h]
\includegraphics[scale=0.8]{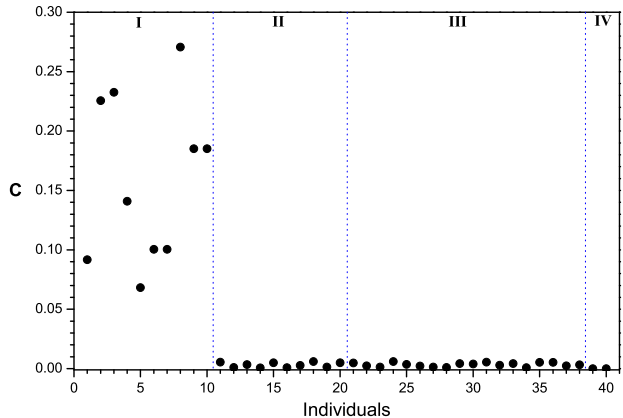}
\caption{Statistical complexity of healthy subjects (group I), 
epileptic patients subject to treatment (group II), epileptic patients without treatment (group III), 
and epileptic patients undergoing seizures (group IV). Vertical lines indicate the limits for each 
group. The order of the individuals in each 
group does not have any meaning.}
\label{meanFieldeegFig}
\end{figure}

Figure~\ref{meanFieldeegFig} shows the statistical complexity of the mean field of the EEG signal for all the individuals in each group from the data base.
Figure~\ref{meanFieldeegFig} indicates that the complexity measure for the global dynamics of the EEG signal 
is higher in healthy subjects (group I) in relation to that of epileptic patients (groups II, III, and IV). 
The lowest levels of complexity correspond
to the patients undergoing epileptic seizures (group IV). 
This measure of complexity allows us to discern between healthy subjects and epileptic 
patients.
Among EEG signals from epileptic patients, those signals from patients 
under treatment seem to possess slightly greater complexity than those from patients without treatment.
However, the quantity $C$ is not very efficient for distinguishing between the different groups of epileptic patients. 
The spread of the values of the complexity found in group I indicates a greater variability in the statistical properties of EEG
signals from the healthy subjects. 

It is important to say 
that other choices for the partition of states of the EEG signal are possible, and hence
different values for the $H$ and $D$  measures can be obtained, without affecting the overall behavior of
the statistical quantity $C$.
For instance, Rosso \textit{et al.} [2003] 
divided the time axis of EEG signals into non-overlapping temporal windows on which
the wavelet energy at different resolution levels can be calculated. 

In order to give an interpretation to the above results, we propose a dynamical model. We consider a system of $N$ 
interacting nonlinear, heterogeneous elements forming a network, where the state of element $i$ ($i=1,2,\dots,N$) at 
time $t$ is denoted by $x_t^i$. The evolution of the state of each element is 
assumed to depend on its own local dynamics and on its interaction with the network,
whose intensity is described by the coupling parameter $\epsilon$. Then, we consider a 
network of maps subjected to a global interaction as follows [Cisneros \emph{et al.}, 2002]
\begin{equation}
\label{systemEq}
x_{t+1}^i = (1-\epsilon)f_i (x_t^i)+ \frac{\epsilon}{N} \sum_{j=1}^N 
f_j(x_t^j),
\end{equation}
where the function $f_i(x_t^i)$ describes the local dynamics of map $i$. 
The usual homogeneous
globally coupled map system [Kaneko, 1990] corresponds to
having the same local function for all the elements, i.e.,
$f_i(x_t^i )=f(x_t^i )$.
As local chaotic dynamics we choose the logarithmic map $f(x)=b+\ln |x|$, 
$x\in (-\infty,\infty)$, where $b$ is a real parameter. This map does not belong to the standard 
class of universality of unimodal or bounded maps. Robust chaos occurs in 
the parameter interval $b\in [-1,1]$, with no periodic windows and no 
separated chaotic bands on this interval [Kawabe \& Kondo, 1991]. Heterogeneity in the maps is
introduced by considering
$f_i(x_t^i)=b_i +\ln |x_t^i|$, 
where the values $b_i$ are uniformly distributed at random in the interval $[-1,1]$. 

As the macroscopic variable for this system, we consider the instantaneous mean field defined as 
\begin{equation}
\label{meanFieldEq}
h_t = \frac{1}{N}\sum_{j=1}^N f_j(x_t^j).
\end{equation}

\begin{figure}[h]
\includegraphics[scale=0.8]{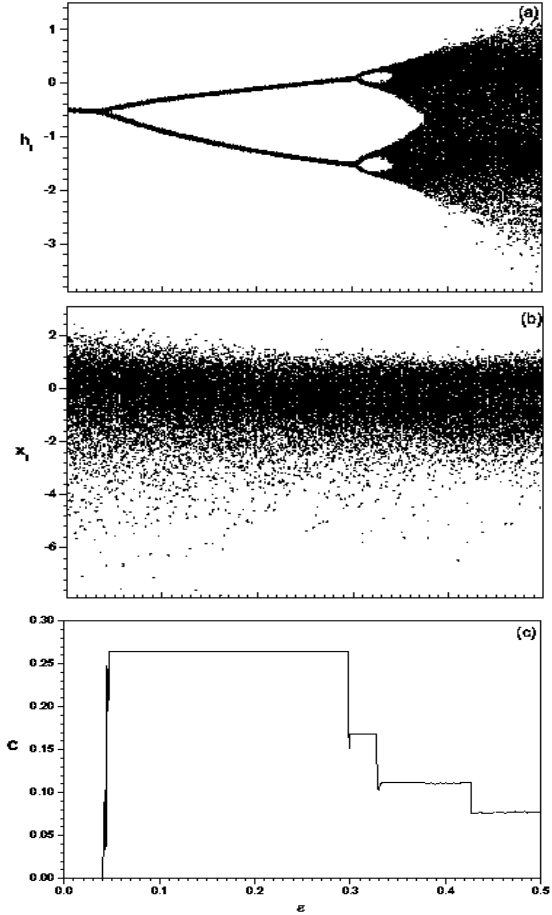}
\caption{(a) Bifurcation diagram of $h_t$ as a function of $\epsilon$ 
for a coupled heterogeneous map network, Eq.~(\ref{meanFieldEq}), with $b_i\in[-1,1]$ and
$N=10^4$. (b) Bifurcation diagram of a local map $x_t^i$ versus $\epsilon$, 
exposing the underlying chaotic dynamics. (c) Complexity $C$  as a function of $\epsilon$.}
\label{systemFig}
\end{figure}

Figure~\ref{systemFig}(a) shows the bifurcation diagram of the mean field 
$h_t$ of the globally coupled heterogeneous map network, Eq.~(\ref{systemEq}), with $b_i \in [-1,1]$ and $N=10^4$,
as a function of the coupling strength $\epsilon$. For each 
value of $\epsilon$, the mean field was calculated at each time step during a 
run of $10^2$ iterates starting from random initial conditions on the local maps, uniformly 
distributed on the interval $x_0^i\in[-8,8]$, after discarding $10^3$ 
transients. The local maps are chaotic
and desynchronized (see Fig.2(b)). However, the mean field
in Fig.~2(a) reveals the existence of global periodic attractors
for some intervals of the coupling parameter. 
This is the phenomenon of nontrivial collective behavior where macroscopic order coexists with local disorder in
a system of interacting dynamical elements. 
Different collective states
emerge as a function of the coupling $\epsilon$. In this representation, collective periodic
states at a given value of the coupling appear as sets
of vertical segments which correspond to intrinsic fluctuations
of the periodic orbits of the mean field. Increasing the
system size $N$ does not decrease the amplitude of the collective
periodic orbits. Moreover, when $N$ is increased the
widths of the segments that make a periodic orbit in the
bifurcation diagrams such as in Fig.~2(a) shrink, indicating
that the global periodic attractors become better defined in
the large system limit.

Figure~\ref{systemFig}(c) shows the complexity $C$ of the mean field as a function of 
$\epsilon$. Here, the observation level was set at $\mathcal{R}=15$.
When the value of $\epsilon$ is small, the bifurcation diagram of $h_t$ shows a period-one collective attractor, 
which implies that for this parameter range 
the system follows the standard
statistical behavior of uncorrelated disordered variables
that yield a single period for its mean field. At the chosen level of resolution, the complexity measure considers 
the macroscopical variable as laying in a single state, thus giving $C=0$. 
The complexity $C$ remains zero up to 
a critical value of the coupling $\epsilon_c \simeq 0.04$. The onset of the complexity at the value $\epsilon_c$ resembles a first order phase transition.
As the periodicity of the collective orbit 
increases, more states are occupied by the probability distribution of the mean field $h_t$.
The probability distribution of $h_t$ corresponding to 
a periodic collective state is not uniform and consists of a set of distinct ``humps''.  A nonuniform probability distribution and few occupied states lead to larger values of the complexity $C$, as observed for the period-two collective orbit.
When the system enters chaotic collective motion, more states are occupied by the probability distribution of the mean field
and therefore this probability becomes more uniform. As a consequence, the complexity decreases.

The emergence of ordered collective behavior in the coupled map network, Eq.~(\ref{systemEq}),
cannot be attributed to the existence of windows of periodicity nor to 
chaotic band splitting in the local dynamics.
Figure~\ref{systemFig} shows that higher values of complexity are associated to the occurrence 
of nontrivial collective behavior in a network of interacting dynamical elements. 
We have obtained similar results for different network topologies and different local map dynamics. 
This result adds support to the concept of complexity as an emergent 
behavior; in this case the ordered collective behavior is not present
at the local level.  
Furthermore, these results suggest that similar dynamical collective processes may take place in 
the human brain. 
From the dynamics of a single neuron as an excitable element it is not possible in general to characterize the collective behavior of the brain, as the collective behavior of the coupled map network cannot be inferred from the knowledge of the dynamics of a single map. 
Thus, the lower values of complexity found in the global dynamics of the 
EEG signals from epileptic patients may be associated to a decrease in the ability to 
generate collective organization and functions in the brain affected by such
physiological condition. 

\begin{figure}[h]
\includegraphics[scale=1]{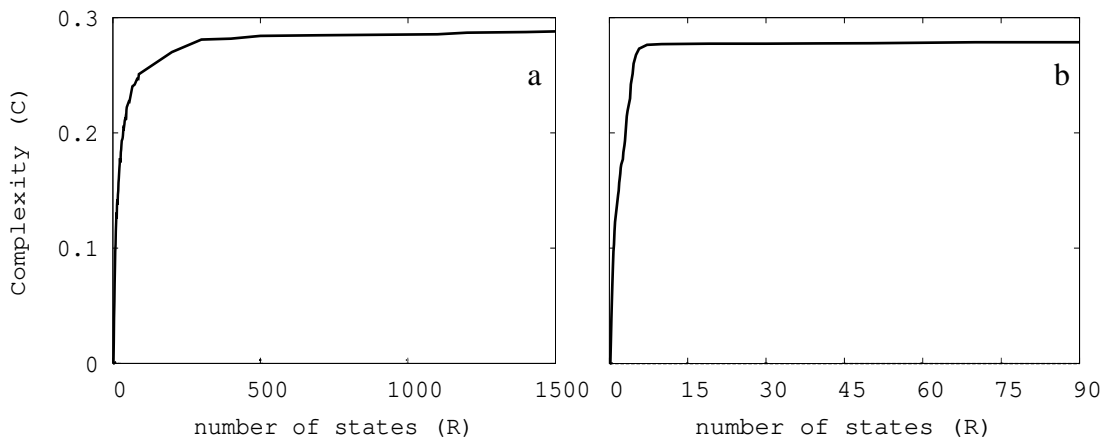}
\caption{(a) Complexity $C$ as a function of $\mathcal{R}$ 
for the mean field from EEG signals for a healthy 
subject. (b) $C$ as a function of $\mathcal{R}$ 
for the coupled heterogeneous map network, Eq.~(\ref{systemEq}), 
$\epsilon=0.2$.}
\label{numberStatesFig} 
\end{figure}

As we have mentioned, the measure of the statistical complexity depends on the 
level of observation (number of states $\mathcal{R}$) considered for the 
computations. Figure~\ref{numberStatesFig}(a) shows the statistical complexity 
as a function of the number of states for the mean field value from an EEG signal of a healthy subject. In Fig.~\ref{numberStatesFig}(b) 
a similar plot of $h_t$ is shown for the value of 
the coupling parameter $\epsilon=0.2$ in the coupled heterogeneous map network, Eq.~(\ref{systemEq}).

Figure~\ref{numberStatesFig} shows that the measures of the complexity in both systems 
tend to constant, asymptotic values when the number of states $\mathcal{R}$ is increased. 
For the coupled map network, the number of states used 
in the computations was $\mathcal{R}=15$, while for the EEG signals we employed
$\mathcal{R}=1000$; these values lie in the corresponding asymptotic regime for each system.

In the heterogeneous coupled maps network, the size of the 
system $N$ can be related to the minimum number of states or level of description $\mathcal{R}_c$ 
that should be used in the calculation of $C$. The width of a vertical segment in the bifurcation diagram of 
$h_t$ at the level of description $\mathcal{R}_c$ is given by $\mathcal{R}_c^{-1}$.  In addition, we have observed that
this width is proportional to the statistical dispersion of the points inside the segment, which in turn
decreases with the size of the system following the law of large numbers as $N^{1/2}$.
Then,
\begin{equation}
\frac{1}{\mathcal{R}_c} \sim \frac{1}{\sqrt N} \Rightarrow N \sim 
\mathcal{R}_c^2,
\end{equation}
thus, for a given system size $N$, it is possible to estimate the 
minimum number of states that should be considered for the computation of the statistical 
complexity in this system. Correspondingly, for the resolution level $\mathcal{R}=15$ that we employed in Fig.~\ref{systemFig},
the minimum system size that could have been used is $N \sim 225$.

In summary, we have shown that the measure of statistical complexity Eq.~(\ref{complexityEq}) used in 
this work is useful in the analysis of theoretical and real systems. There exists
a morphological difference between the global dynamics in EEG signals from healthy subjects 
and those from epileptic patients that can be revealed by employing this tool. 
The calculations show that epilepsy is associated to a low level of statistical complexity. 

We have also shown that the mean field of a network of heterogeneous chaotic maps contains 
relevant information about the complexity of the system which cannot be derived from the knowledge of the
behavior of the local maps. A high degree of complexity is associated to the emergence of 
nontrivial collective behavior.
These results can be related to those of Cisneros \emph{et al.} [2002] 
who showed that the prediction error used to measure the mutual information transfer between a local and a 
global variable in a similar network decreased when nontrivial collective behavior arises in the system. 
Thus, a high level of complexity can also be associated to an increase in the mutual information 
transfer between macroscopic and microscopic variables in a system. 

Finally, our intention is not to describe the human as network of interacting chaotic elements, but to show
that the collective properties related to the concept of complexity are 
qualitatively similar, and that a rise in complexity can be associated with 
the emergency of collective ordered behavior in different systems.

\section*{Acknowledgments}
M. G. C. acknowledges support from Consejo de Desarrollo Cient\'{\i}fico, Human\'{\i}stico 
y Tecnol\'ogico (C.D.C.H.T.), Universidad de Los Andes (ULA), Venezuela, under grant C-1579-08-05-B.  M.E-M acknowledges support from C.D.C.H.T.-ULA under grant I-1075-07-02-B. 
R.L-R. acknowledges support from Postgrado en F\'{\i}sica Fundamental, 
Universidad de Los Andes, Venezuela, during his visit there in July 2008. 
P. G. thanks C.D.C.H.T, Universidad Central de Venezuela, for support. 
The authors thank Hospital Luis Gomez Lopez, in Barquisimeto, Venezuela, for providing
the clinical data base and diagnoses. 

\newpage

\noindent {\bf References} \smallskip

\noindent Badii, R. \&  Politi, A. [1997] \textit{Complexity: Hierarchical structures 
and scaling in Physics}, Cambridge University Press.  

\noindent Babloyantz A. \& Destexhe A. [1986] ``Low-dimensional chaos in an instance of
epilepsy'', \textit{Proc. Natl. Acad. Sci. U. S. A.} {\bf 83}, 3513-3517.  

\noindent Chat\'e, H. \& Manneville, P. [1992a] ``Collective behaviors in spatially extended systems with local interactions and synchronous updating'', \textit{Prog. Theor. Phys.} {\bf 87}, 1-60.

\noindent Chat\'e, H. \& Manneville, P.  [1992b] ``Emergence of effective low-dimensional dynamics in the macroscopic behaviour of coupled map lattices'', \textit{Europhys. Lett.} {\bf 17}, 291-296.

\noindent Cisneros, L., Jim\'enez, J.,  Cosenza, M. G. \&  
Parravano, A. [2002] ``Information transfer and nontrivial collective behavior in chaotic coupled map networks'',  \textit{Phys. Rev. E} {\bf 65}, 045204R.

\noindent Cosenza, M. G. [1998] ``Nontrivial collective behavior in coupled maps on fractal lattices'', \textit{Physica A} {\bf 257}, 357-364.

\noindent Cosenza, M. G. \& Gonz\'alez, J. [1998] ``Synchronization and collective behavior in globally coupled logarithmic maps'', \textit{Prog. Theor. Phys.} {\bf 100}, 21-38.

\noindent Herz, A. V. M., Gollisch, T.,  Machens, C. K. \& Jaeger, D. [2006] ``Modeling single-neuron dynamics and computations: A balance of detail and abstraction'', \textit{Science} \textbf{314}, 80-88.

\noindent Jasper, H. H. [1958] ``Report of the committee on methods of clinical examination in electroencephalography'', \textit{Electroenceph. Clin. Neurophysiol.} {\bf 10}, 370-375.

\noindent Kaneko, K. [1990] ``Globally coupled chaos violates the law of large numbers but not the central-limit theorem'', \textit{Phys. Rev. Lett.} {\bf 65}, 1391-1394.

\noindent Kaneko, K. \&  Tsuda, I. [2000], \textit{Complex Systems: Chaos and beyond}, Springer, Berlin.

\noindent Kawabe, T. \&  Kondo, Y. [1991] ``Fractal transformation of the one-dimensional chaos produced by logarithmic map'', \textit{Prog. Theor. Phys.} {\bf 85}, 759-769.

\noindent L\'opez-Ruiz, R.,  Mancini, H. L. \&  Calbet, X. [1995] ``A statistical measure of complexity'', \textit{Phys. Lett. 
A} {\bf 209}, 321-326.

\noindent Manrubia, S. C.,  Mikhailov, A. \& Zanette, D. H. [2004], \textit{Emergence of Dynamical Order}, 
World Scientifc, Singapore.

\noindent Martinerie, J., Adam, C., Le Van Quyen, M., Baulac, M., Cl\'emenceau, S., Renault, B. \& Varela, F. [1998] ``Epileptic seizures can be anticipated by non-linear analysis'', 
\textit{Nat. Med.} {\bf 4}, 1173-1176.

\noindent Mikhailov, A. \& Calenbuhr, V. [2002], \textit{From Swarms to Societies: Models of complex behavior}, 
Springer, Berlin.

\noindent Pikovsky, A. S. \&   Kurths, J. [1994] ``Do globally coupled maps really violate the law of large numbers?'', \textit{Phys. Rev. Lett.} {\bf 72}, 1644-1646.

\noindent Rapp, P. E.,  Zimmerman, I. D., Vining, E. P., Cohen, N., Albano, A. M. \& Jimenez-Montano, M. A. [1994] ``The algorithmic complexity of neural spike trains increases during focal seizures'', \textit{J. Neuroscience} {\bf 14}, 4731-4739.

\noindent Rosso, O. A., Martin, M. T. \& Plastino, A. [2003] ``Brain electrical activity analysis using wavelet based informational tools (II): Tsallis non-extensivity and complexity measures'', \textit{Physica A} {\bf 320}, 497-511. 

\noindent S\'anchez, J. R. \&  L\'opez-Ruiz, R. [2005] ``A method to discern complexity in two-dimensional patterns
generated by coupled map lattices'', \textit{Physica A} {\bf 355}, 633-640.

\noindent Shibata, T. \& Kaneko, K. [1998] ``Collective chaos'', \textit{Phys. Rev. Lett.} {\bf 81}, 4116-4119.

\end{document}